\documentclass{ws-ijmpe}

\begin{document}

\markboth{A. K. Chaudhuri}{Dissipative hydrodynamics in 2+1 dimensions }

\catchline{}{}{}{}{}
%%%%%%%%%%%%%%%%%%%%%%%%%%%%%%%%%%%%%%%%%%%%%%%%%%%%%%%%%%%%%%%%%%%%

\title{Dissipative hydrodynamics in 2+1 dimensions}

\author{\footnotesize A. K. Chaudhuri}

\address{Variable Energy Cyclotron Centre\\1-AF, Bidhan Nagar, Kolkata 700064\\ 
akc@veccal.ernet.in}

\maketitle

\begin{history}
\received{(received date)}
\revised{(revised date)}
%\accepted{(Day Month Year)}
%\comby{(xxxxxxxxxx)}
\end{history}

\begin{abstract}
In a first order theory of dissipative hydrodynamics, we have simulated hydrodynamic
evolution of QGP fluid  with dissipation due to shear viscosity only.  Simulation confirms that compared to an ideal fluid,   energy density or temperature of a viscous fluid evolve slowly. Transverse expansion is also more in viscous fluid. We also study the effect of viscosity on particle production. Particle production is enhanced, more at large $p_T$. The elliptic flow on the otherhand decreases and shows a tendency to saturate at large $p_T$.
\end{abstract}

\section{Introduction}
 
A large volume of experimental data from Au+Au collisions at RHIC are successfully analysed in an {\em ideal} hydrodynamic model \cite{QGP3}.   However, experimental data do show deviation from ideal behavior.  The ideal
fluid description works well in almost central Au+Au collisions near  mid-rapidity at top RHIC energy, but gradually breaks
down in more  peripheral collisions, at forward rapidity, or at lower collision energies, indicating the
onset of dissipative effects. To describe such deviations from ideal fluid dynamics quantitatively,  requires the  numerical
implementation of {\em dissipative} hydrodynamics.

Though the theories of dissipative hydrodynamics \cite{Eckart,IS79} 
has been known for more than 30 years , significant progress toward its numerical implementation  has only been made very
recently 
\cite{Teaney:2004qa,Muronga:2001zk,MR04,Chaudhuri:2005ea,Heinz:2005bw,Chaudhuri:2006jd}.
At the Variable Energy Cyclotron Centre, Kolkata, we have developed a  numerical code  (AZHYDRO-KOLKATA) to solve,   the 1st
order   dissipative hydrodynamics in 2+1 dimension (assuming boost-invariance in the longitudinal direction) and currently extending the code to 2nd order dissipative hydrodynamics. Some of the results from 1st order viscous hydrodynamics will be presented here.

\section{1st order dissipative fluid dynamics}
 
Relativistic dissipative hydrodynamics has been discussed in detail in ref.\cite{Chaudhuri:2005ea,Heinz:2005bw,Chaudhuri:2006jd}. 
In dissipative hydrodynamics, dissipative fluxes are assumed to be small. The entropy current is expanded in terms of dissipative fluxes. In the 1st order theory \cite{Eckart}, the expansion contain terms linear in dissipative fluxes, explicit form of which could be obtained by satisfying the second law of thermodynamics $\partial_\mu S^\mu \geq 0$. 
1st order theories are acausal, signal can travel faster than light. This is corrected in 2nd order theory \cite{IS79}, entropy expansion contains terms quadratic in dissipative fluxes. Naturally, 2nd order theories are more complex. 

We consider a QGP fluid in the central rapidity region with net zero baryon density ($n_B=0$). We also neglect all the dissipative effects (e.g. heat conduction   and bulk viscosity) other than the shear viscosity. We work in the Landau energy frame. Energy-momentum tensor, including the shear pressure tensor $\pi^{\mu\nu}$ is written as,

\begin{equation}
\label{eq4}
 T^{\mu\nu}  
           = \varepsilon\,u^\mu u^\nu -p\Delta^{\mu\nu}
               + \pi^{\mu\nu} \\
 \end{equation}

 \noindent where  $u^\mu$ is the hydrodynamic 4-velocity ($u^\mu u_\mu=1$)  and  $\Delta^{\mu\nu}=g^{\mu\nu}-u^\mu u^\nu$ is the projector, orthogonal to $u^\mu$.
$\varepsilon=u_\mu T^{\mu\nu} u_\nu$ is the energy density, $p$ is the hydrostatic  
pressure. $T^{\mu\nu}$ satisfies the conservation law,

\begin{equation}
\partial_\mu T^{\mu\nu}=0
\end{equation}

  In the first order theories \cite{Eckart,IS79} , the shear stress tensors are written as,
 
\begin{equation}
\label{eq6}
\pi^{\mu\nu} = 2\eta  \nabla^{<\mu}u^{\nu>}=2\eta
\left[\frac{1}{2}\left(\Delta^{\mu\sigma}
\Delta^{\nu\tau}{+}\Delta^{\nu\sigma}\Delta^{\mu\tau}\right)-\frac{1}{3}
\Delta^{\mu\nu}\Delta^{\sigma\tau}\right] 
\end{equation}

\noindent where $\eta$ is the shear viscosity coefficient.  
 $\pi^{\mu\nu}$  is symmetric ($\pi^{\mu\nu}=\pi^{\nu\mu}$),
traceless ($\pi^\mu_\mu=0$) and transverse to hydrodynamic velocity,
($u_\mu \pi^{\mu\nu}=0)$. The 16-component $\pi^{\mu\nu}$ has only 5
independent components. 

Heavy ion collisions are best described in terms of proper time $\tau=\sqrt{t^2-z^2}$ and
rapidity $\eta_s=\frac{1}{2}\ln \frac{t+z}{t-z}$.
In $(\tau,x,y,\eta_s)$ coordinates, with longitudinal boost-invariance,
the hydrodynamic 4-velocity can be written as, 

\begin{eqnarray}
u^\mu=(u^\tau,u^x,u^y,u^{\eta_s}) 
          =(\gamma_\perp,\gamma_\perp v_x, \gamma_\perp v_y,0),
\end{eqnarray}

%$\gamma_\perp(1,v_x,v_y,0)$, with
\noindent with $\gamma_\perp=1/\sqrt{1-v_x^2-v_y^2}$.
 The 
energy-momentum conservation equations are,

%\begin{widetext}
%\label{eq8}
\begin{eqnarray}
\label{eq8a}
&& \partial_\tau \tilde{T}^{\tau\tau}  
 +\partial_x (\tilde{T}^{\tau \tau} \overline{v}_x  )+\partial_y  (\tilde{T}^{\tau \tau} \overline{v}_y)= 
  -\,(p+\tau^2 \pi^{\eta\eta})
\\
\label{eq8b}
&&\partial_\tau\tilde{T}^{\tau x} 
 +\partial_x (\tilde{T}^{\tau x}v_x)
 +\partial_y (\tilde{T}^{\tau x}v_y) 
 = -\partial_x(\tilde{p} + \tilde{\pi}^{xx}-\tilde{\pi}^{\tau x} v_x) - \partial_y(\tilde{\pi}^{xy}-\tilde{\pi}^{\tau x}v_y)
\\
\label{eq8c}
 &&\partial_\tau\tilde{T}^{\tau y} 
 +\partial_x (\tilde{T}^{\tau y}v_x)
 +\partial_y (\tilde{T}^{\tau y}v_y) 
 = -\partial_x(\tilde{\pi}^{xy}-\tilde{\pi}^{\tau y} v_x) - \partial_y(\tilde{p} + \tilde{\pi}^{yy}-\tilde{\pi}^{\tau y}v_y)
\end{eqnarray}

\noindent where $\overline{v}_x=T^{\tau x}/T^{\tau\tau}$ and $\overline{v}_y=T^{\tau y}/T^{\tau\tau}$, and we have used
the notation
"tilde" to represent quantities multiplied by the factor $\tau$, 
$\tilde{p}=\tau p$ and similarly   $\tilde{T}^{ij}=\tau T^{ij}$. 
We note that unlike in ideal fluid, in viscous fluid dynamics 
conservation equations contain additional pressure gradients containing
the dissipative fluxes.  
Both $T^{\tau x}$ and $T^{\tau y}$ components of energy-momentum
tensors now evolve under
the influence of additional pressure gradients.  

With boost-invariance, number of independent shear stress tensor reduces to 3. $\pi^{\tau\tau}$, $\pi^{xx}$ and $\pi^{yy}$
can be chosen as independent components (there could be other choices also \cite{Heinz:2005bw}),

\begin{eqnarray} \label{9}
 &&\pi^{\tau\tau} =2\eta \left [ \frac{\theta}{3}(\gamma_\perp^2-1) 
  + \partial_\tau\gamma_\perp-\frac{1}{2}D(\gamma_\perp^2) \right ],
\\
&&\pi^{xx}=2\eta \left [-\partial_x(\gamma_\perp v_x)-\frac{1}{2}D(\gamma_\perp^2 v_x^2)+
 \frac{\theta}{3}(1+\gamma_\perp^2 v_x^2) \right ], \\
&&\pi^{yy}=2\eta \left [-\partial_y(\gamma_\perp v_y)-\frac{1}{2}D(\gamma_\perp^2 v_y^2)+
 \frac{\theta}{3}(1+\gamma_\perp^2 v_y^2) \right ]
\end{eqnarray}
%\end{widetext}

\noindent where $D=u^\mu \partial_\mu
=\gamma_\perp(\partial_\tau + v_x \partial_x + v_y \partial_y)$
is the convective time derivative,
and $\theta$ is the  local expansion rate, 
$\theta=\frac{\gamma_\perp}{\tau}+\partial_\tau \gamma_\perp +
\partial_x(v_x\gamma_\perp)+\partial_y(v_y \gamma_\perp)$.
 
The other shear stress-tensors required in solving Eqs.\ref{eq8a},\ref{eq8b},\ref{eq8c}  can be obtained from the constraints satisfied by $\pi^{\mu\nu}$, (i) $\pi^\mu_\mu=0$:tracelessness, (ii)  $u^\mu \pi_{\mu\nu}=0$:  
transverse to $u^\mu$.

\begin{eqnarray}
\tau^2\pi^{\eta\eta}=&&\pi^{\tau\tau}-\pi^{xx}-\pi^{yy} \\
2v_x \pi^{x\tau}=&&\pi^{\tau\tau}+v_x^2\pi^{xx}-v_y^2\pi^{yy}\\
2v_y \pi^{y\tau}=&&\pi^{\tau\tau}-v_x^2\pi^{xx}+v_y^2\pi^{yy}\\
2v_xv_y \pi^{xy}=&&\pi^{\tau\tau}-v_x^2\pi^{xx}-v_y^2\pi^{yy}
\end{eqnarray}

Shear stress-tensor components contains 
time derivatives of velocities $v_x$ and $v_y$. 
  Thus at time step $\tau_i$ one needs the  
still unknown time derivatives.  
In 1st order theories, this problem is circumvented by calculating the
time derivatives from the ideal equation of motion ,
$Du^\mu=\frac{\nabla^\mu p}{\varepsilon+p}$ and
$D\varepsilon=-(\varepsilon+p)\nabla_\mu u^\mu$.

\section{Equation of state, viscosity coefficient and initial conditions}

Through the equation of state, the macroscopic hydrodynamic models make contact with the microscopic world. 
We have used the equation of state,
EOS-Q, developed in ref.\cite{QGP3}.  It is a two-phase equation of
state. The hadronic phase of EOS-Q is modeled
as a non-interacting gas of hadronic resonance.   The
QGP phase  is modeled as that of a 
non-interacting quarks (u,d and s) and gluons, confined by a bag pressure
B.   Adjusting the Bag pressure, the two phases 
are matched by Maxwell construction at the critical temperature, $T_c=164 MeV$.

Shear viscosity coefficient ($\eta$) of dense nuclear (QGP or 
resonance hadron gas) is quite uncertain.
In perturbative regime, shear viscosity of a QGP is
estimated \cite{Arnold:2000dr}, $\eta=86.473 \frac{1}{g^4}\frac{T^3}{log g^{-1}}$.  
With entropy of QGP, $s=37\frac{\pi^2}{15}T^3$ and 
$\alpha_s \approx$0.5, the ratio of viscosity over the entropy, in the perturbative regime is estimated as,

\begin{equation}
\left (\frac{\eta}{s} \right )_{pert} \approx 0.135,
\end{equation}

However, QGP  produced in nuclear collisions is non-perturbative. It is
 strongly interacting QGP. Recently, using the ADS/CFT correspondence
\cite{Policastro:2001yc},
shear viscosity of a strongly coupled gauze theory, N=4 SUSY YM,
has been evaluated, $\eta=\frac{\pi}{8}N^2_cT^3$ and the    entropy
is given by $s=\frac{\pi^2}{2}N^2_cT^3$. Thus in the strongly coupled field theory,

\begin{equation}
\left ( \frac{\eta}{s} \right )_{ADS/CFT} = \frac{1}{4\pi}\approx0.08,
\end{equation}

To demonstrate the effect of viscosity on flow and subsequent particle
 production, we use both the perturbative and ADS/CFT estimate of viscosity.

Solving Eqs.\ref{eq8a},\ref{eq8b},\ref{eq8c} require initial conditions.
In the present demonstrative calculations,  we have used the 
similar initial conditions as in ref.\cite{QGP3}. 
  We just mention that
in ref.\cite{QGP3}, initial
transverse energy is parameterised geometrically, with 25\% hard scattering. At initial time $\tau_i$=0.6 fm, the central entropy density is $s_{ini}$=110  $fm^{-3}$. Fluid velocities are assumed to be zero initially.
In dissipative hydrodynamics, dissipative fluxes need to be specified also. 
We assume that by the equilibration time $\tau_i$, the dissipative fluxes
attained their longitudinal  boost-invariant values. The independent  components at initial time are, $\pi^{xx}=2\eta/\tau_i$,  
$\pi^{yy}=2\eta/\tau_i$ and $\pi^{\tau\tau}=0$. 
 
\vspace{-0.5cm}
\begin{figure}[h]
\begin{minipage} {14pc} 
\includegraphics[width=14pc]{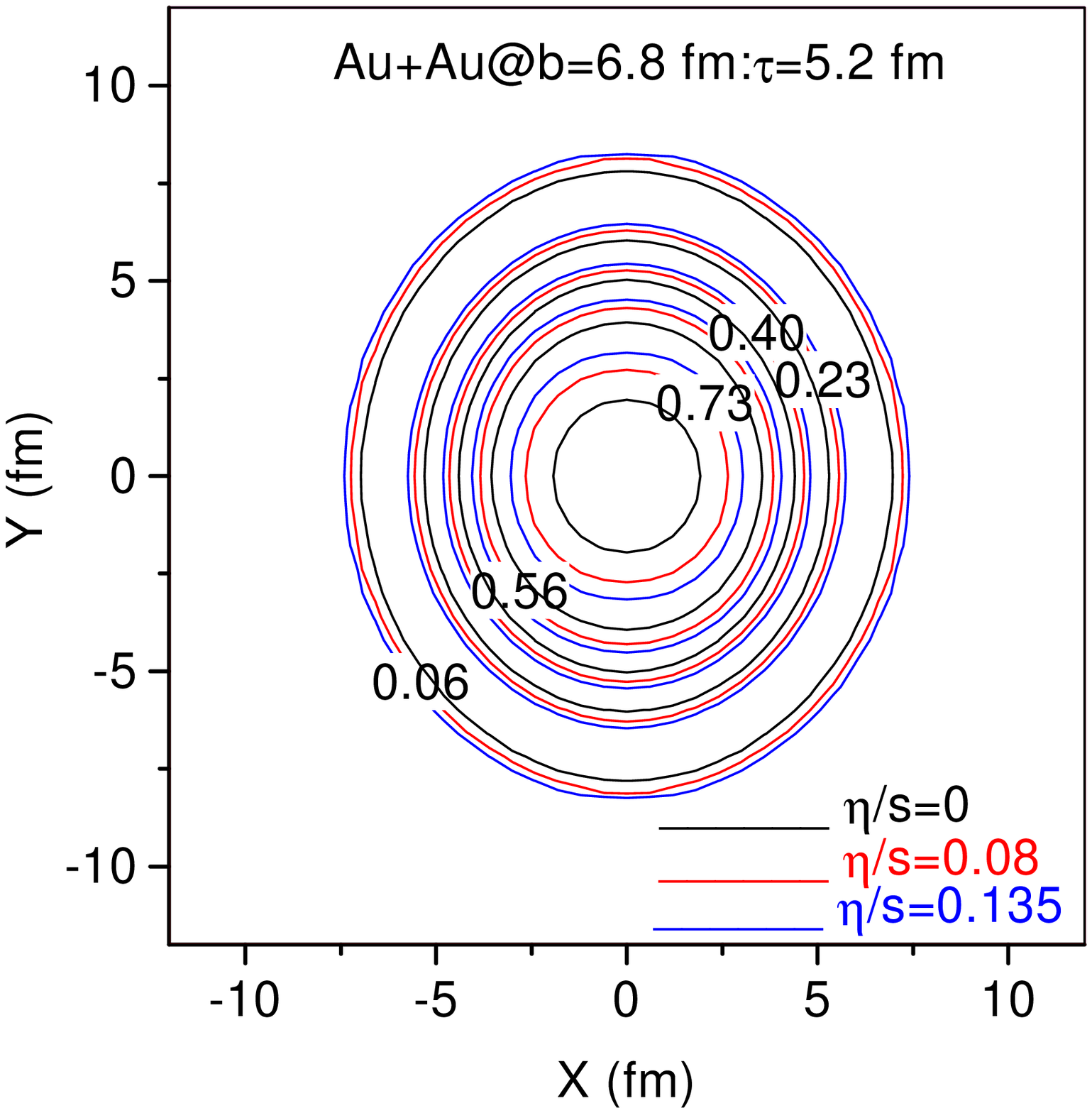}
\vspace{-3.5cm}
\caption{\label{F1}
Contours of constant local energy density in the $x$-$y$ plane 
  at   $\tau$5.2 fm.}
\end{minipage}  \hspace{1pc}%
\begin{minipage} {14pc}
\includegraphics[width=14pc]{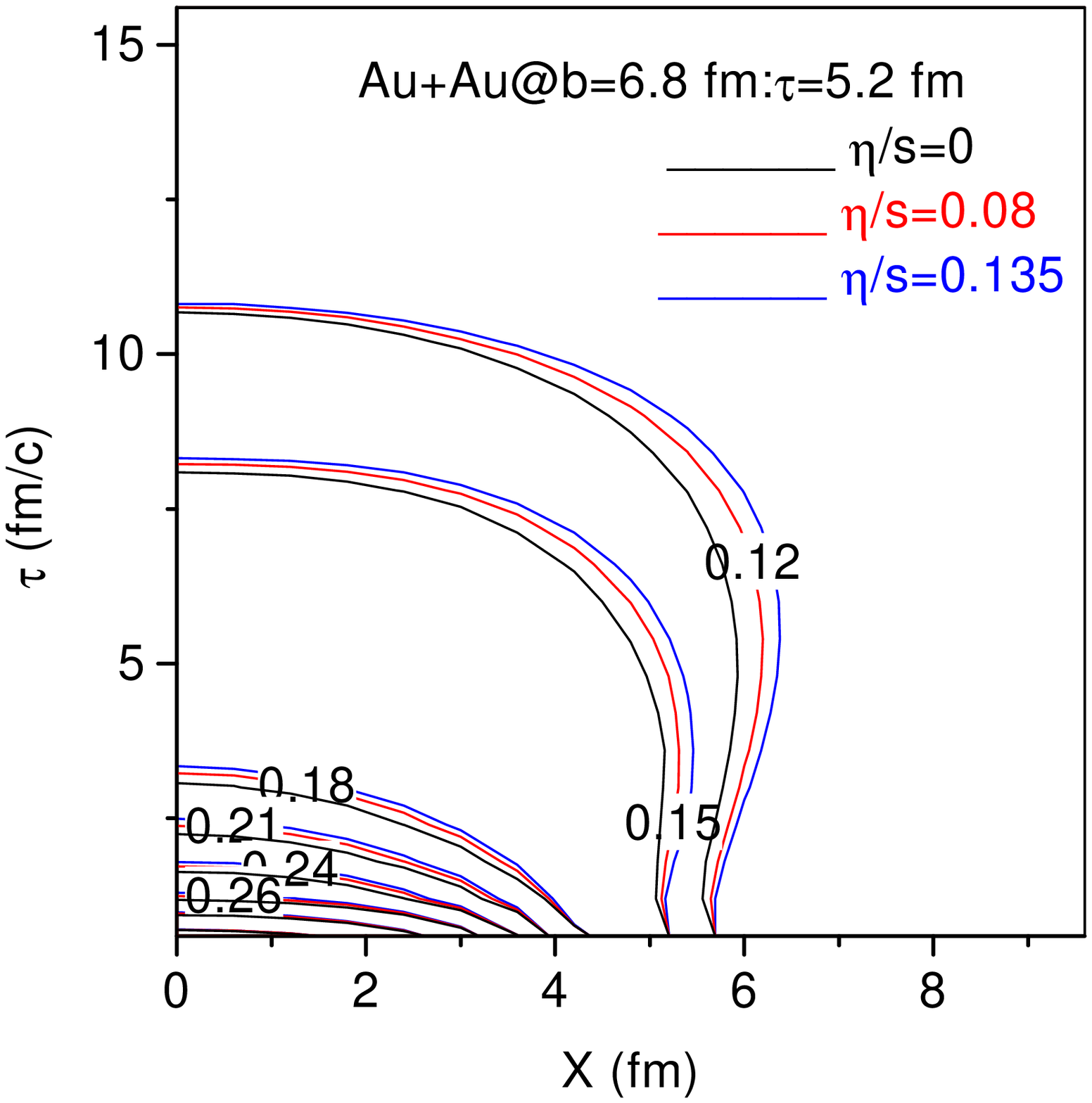}
\vspace{-3.5cm}
\caption{\label{F2}
Contours of constant temperature in $\tau$-x plane. y=0. 
}
\end{minipage} 
\end{figure}
%%%%%%%%%%%%%%%%%%%%%%%%%%%%%%%%%%%%%%%%%%%

 \section{Evolution of the viscous fluid}
To demonstrate the effect of viscosity,
  with the same initial conditions,
we have solved the energy-momentum conservation equations for
ideal fluid and viscous fluid.  

%%%%%%%%%%%%%%%%%%%%%%%%%%%%%%%%%%%%%%%%%%% 
%%%%%%%%%%%%%

 In Fig.\ref{F1}, we have shown
the constant energy density contour plot in x-y plane, after an evolution of 
5.2 fm.
The black lines are for ideal fluid evolution. The  red and blue lines are for
viscous fluid with ADS/CFT ($\eta/s$=0.08) and perturbative ($\eta/s$=0.135) estimate of viscosity. With viscosity fluid cools slowly. Cooling
gets slower as viscosity increases.
To obtain an idea of transverse expansion of viscous fluid, as opposed
to ideal fluid, in Fig.\ref{F2}, we have shown the constant temperature
contours in $\tau-x$ plane ,  at a fixed value of y=0 fm. Transverse expansion
is substantially enhanced in a viscous fluid. More the viscosity, more
is the transverse expansion. The plot also indicate
that at late time, fluid at x=y=0 behaves similarly to a ideal fluid.  

1st order theories are acausal, signal can travel faster than light. Acausality can lead to  unphysical effects like reheating early in the collisions. However, we have not found any evidence of reheating. For small viscosity, with initial conditions as required in RHIC energy Au+Au collisions, effect of causality violation is minimum.

%%%%%%%%%%%%%%%%%%%%%%%%%%%%%%%%%%%%%%%%%%%%%%%%%%%%%%%%%%%%%% 
\vspace{-0.5cm}
\begin{figure}[h]
\begin{minipage} {14pc} 
\includegraphics[width=14pc]{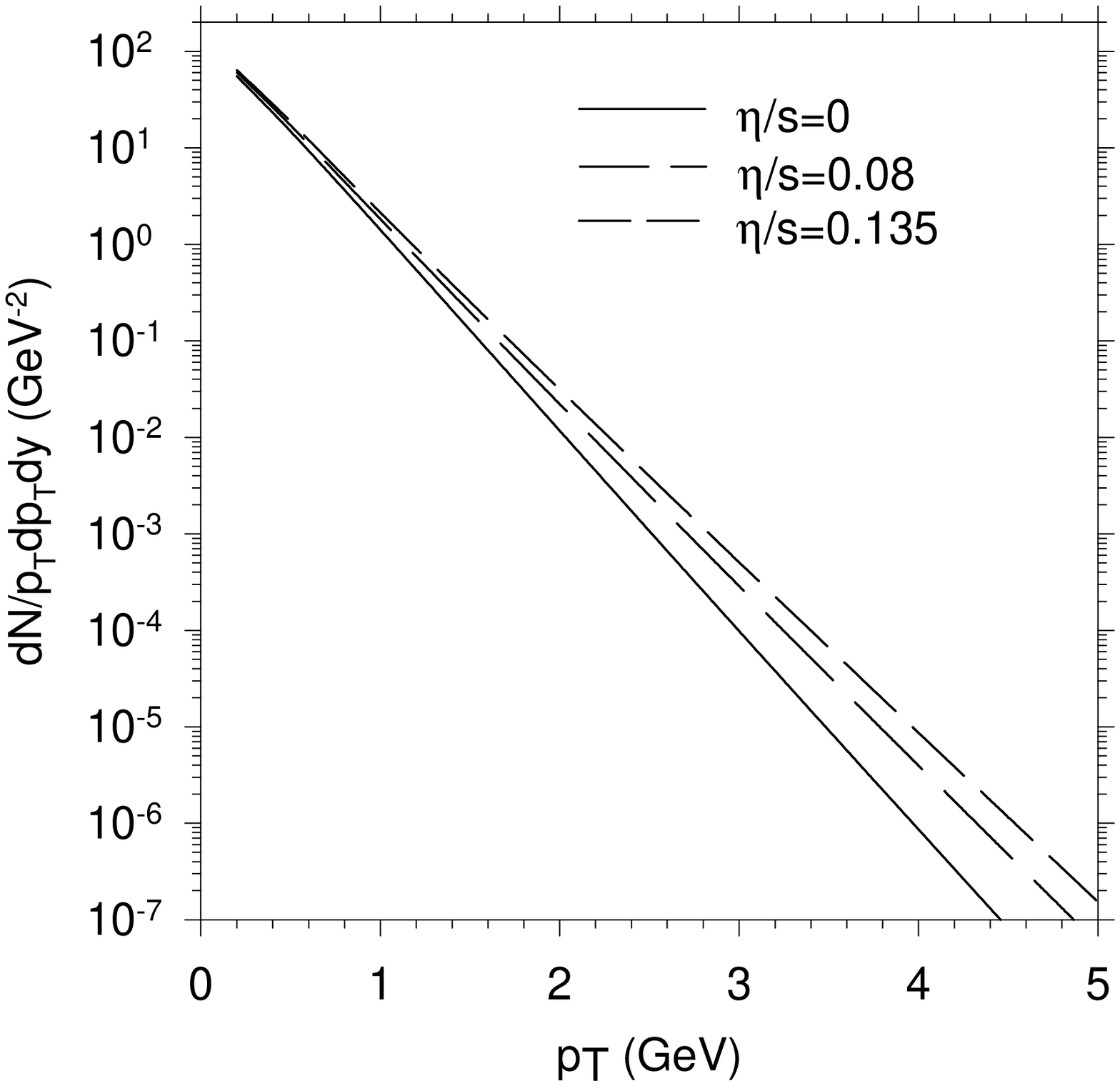}
\vspace{-3.5cm}
\caption{\label{F3}
$P_T$ distribution of $\pi^-$ for ideal and viscous fluid. 
 }
\end{minipage}  \hspace{1pc}%
\begin{minipage} {14pc}
\includegraphics[width=14pc]{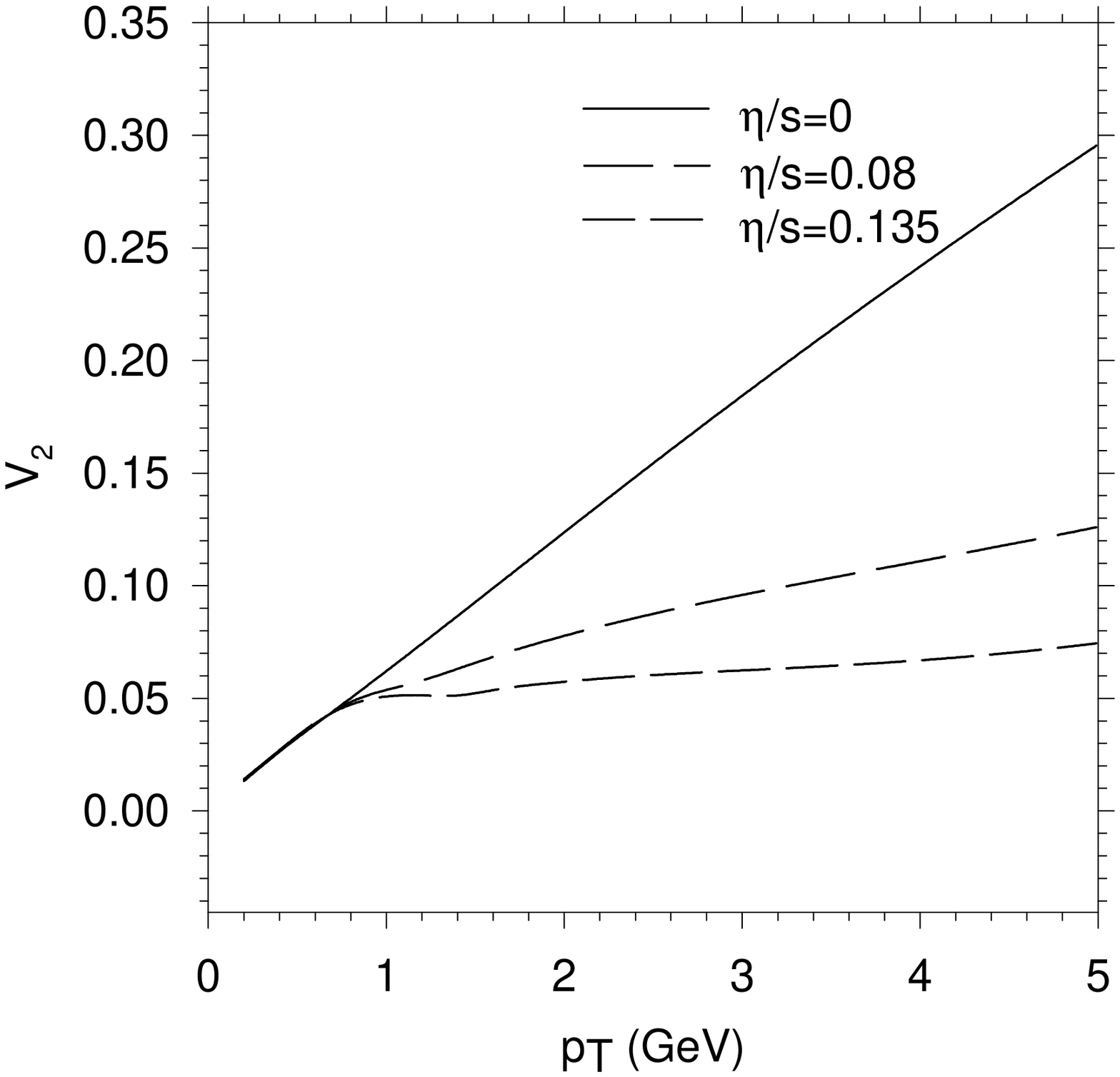}
\vspace{-3.5cm}
\caption{\label{F4}
Elliptic flow as a function of the transverse momentum. 
}
\end{minipage} 
\end{figure}
%%%%%%%%%%%%%%%%%%%%%%%%%%%%%%%%%%%%%%%%%%%%%%%%%%%%%%%%
 
\section{Particle spectra and elliptic flow}
Viscosity generates entropy and particle production is enhanced.
Viscosity
influences the particle production by (i) changing the freeze-out surface (freeze-out surface is extended) and (ii) by introducing
a correction to the
equilibrium distribution function. For small departure from the equilibrium,  
the non-equilibrium distribution function can be  approximated  as, 

\begin{equation} \label{eq17}
f(x,p)=f^{(0)}(x,p) [1+\phi(x,p)],
\end{equation}

\noindent  $\phi(x,p)$ is the deviation from equilibrium distribution 
function  $f^{(0)}$. With shear viscosity as the only dissipative force,
$\phi(x,p)$ can be locally approximated as, 
\begin{equation}\label{eq18}
\phi(x,p)=C\pi_{\mu\nu} p^\mu p^\nu; \hspace{.5cm} C=\frac{1}{2T^2(\varepsilon+p)}
\end{equation}

\noindent completely specifying the non-equilibrium distribution function. The standard 
Cooper-Frye prescription for particle production from freeze-out surface ($\Sigma_\mu$) can be employed to obtain the particle spectra, Detailed expressions for $EdN/d^3p$ with non-local equilibrium distribution function Eqs.\ref{eq17},\ref{eq18} can be found in ref. \cite{Chaudhuri:2006jd}.

%\begin{equation} \label{eq5_1}
%E\frac{dN}{d^3p}=\frac{dN}{dyd^2p_T} =\int_\Sigma %d\Sigma_\mu p^\mu f(x,p)
%\end{equation}

In Fig.\ref{F3}, $p_T$ spectra for $\pi^-$, from freeze-out surface at $T_F=158 MeV$ is shown. Particle  production is increased in viscous dynamics.   We also note that effect of viscosity
is more prominent at large $p_T$ than at low $p_T$. $p_T$ spectra of pions are flattened with viscosity.  

We have also calculated the elliptic flow in the model. Being a ratio, 
elliptic flow is very sensitive to the model. Experimentally, elliptic flow
saturates at large $p_T$. 
In Fig.\ref{F4} $p_T$ dependence of elliptic flow is shown.  
 Elliptic flow decreases with viscosity. As viscosity increases, elliptic flow
  reduces. We also note that both for ADS/CFT and perturbative estimate of viscosity, elliptic flow indicate
saturation at large $p_T$.
The result is very encouraging, as experimentally also elliptic flow
tends to saturate at large $p_T$.
 
 \section{Summary and conclusions}

In a 1st order theory of dissipative hydrodynamics, we have studied the boost-invariant hydrodynamic evolution of 
QGP fluid with dissipation due to shear viscosity.  
In this model study, we have 
considered two values of viscosity, the ADS/CFT motivated value,
$\eta/s \approx$0.08 and perturbatively estimated viscosity, $\eta/s \approx$0.135.
Both the ideal and viscous fluids are initialised similarly.  
 Explicit simulation of ideal and viscous fluids confirms that energy density/temperature of a 
viscous fluid evolves slowly than its ideal counterpart.  Transverse expansion is also more in viscous dynamics.
For a similar freeze-out condition freeze-out surface is extended in
viscous fluid.

We have also studied the effect of viscosity on particle production. Due to viscosity particle production is enhanced, more at large $p_T$. The elliptic flow on the otherhand reduces
and shows a tendency of saturation at large $p_T$.

%%%%%%%%%%%%%%%%%%%%% Bibliography %%%%%%%%%%%%%%%%%%%%%%%%%%%%%%%%%%%%%%%%

  \end{document}